\documentclass[preprint,12pt]{elsarticle}

\usepackage{amssymb}
\usepackage[fleqn]{amsmath}
\usepackage{float}
\usepackage{xcolor}

\begin{document}

\begin{frontmatter}



\title{Modal and wave synchronization in coupled self-excited oscillators.} 


\author{Y. Wolfovich, O.V. Gendelman} 

\affiliation{organization={Faculty of Mechanical Engineering, Technion–Israel Institute of Technology},
            city={Haifa},
            postcode={3200003}, 
            country={Israel}}

\begin{abstract}
In addition to a common synchronization and/or localization behavior, a system of linearly coupled bistable similar Van der Pol (BVdP) oscillators can exhibit a "non-conventiional" or "modal" synchronization. In two-DOF case, one can observe stable beatings attractor with synchronized amplitudes of the symmetric and antisymmetric modes. Current study demonstrates that this unusual behavior is generic and quite ubiquitous, if the system is explored in appropriate parametric regime. Indeed, in the absence of the self - excitation the system of coupled linear oscillators possesses a complete set of non-interacting eigenmodes. If the system is symmetric and the coupling is weak enough, appropriate initial conditions will result in a continuous multi-parametric family of stationary beatings or beat waves. If the self-excitation terms are small enough, i.e. even weaker than the weak coupling, one can expect that, generically, some special stationary or wave beatings will turn into the stable attractors. We demonstrate this phenomenon in systems of ring-coupled BVdP oscillators. In particular, we observe simple two-wave synchronization for $N=2,3,5,6,7$ coupled oscillators; for $N=2$ it corresponds to the "modal" synchronization observed previously. The case $N=4$ is special: due to internal resonances in the slow flow, the two-wave synchronization turns unstable, and more complicated patterns of the multi-wave synchroniztion are revealed. Analytic models manage to capture the shapes of the observed beat waves. All results are verified by direct numeric simulations. 
\end{abstract}

\begin{highlights}
\item Phenomenon of modal and wave synchronizatio in coupled self-excited oscillators is presented
\item The observations and underlying theory explain earlier findings on non-conventional synchronization.
\item Generic and ubiquitous nature of the phenomenon exhibits itself in ana appropriate paremetric regime
\end{highlights}


\end{frontmatter}



\section{Introduction}
The Van der Pol (VdP) oscillator was introduced to scientific literature in the beginning of the 20th century \cite{van1920theory} and has since become a key model for systems with a limit cycle (LC) as a single attractor. In biophysics, it has been used to investigate a wide range of phenomena, including heartbeats \cite{vdpvandermark1928}, Parkinsonian tremors \cite{beuter2003data}, eye chemistry \cite{rompala2007dynamics, camacho2004dynamics}, EEG dynamics \cite{kirschfeld2005physical}, vocal fold oscillations during phonation \cite{garrel2008using, lucero2011lumped}, and neuronal action potentials \cite{fitzhugh1961impulses}. Its applications also span fields like radio electronics, seismology, nonlinear optics, and beyond. Variants of the VdP oscillator can include higher-order terms for the nonlinear damping. A bi-stable version of the VdP oscillator can be easily formulated with two attractors: e.g., a stable fixed point and a stable limit cycle, separated by an unstable limit cycle \cite{defontaines1990chain}. Studies on discrete chains of bi-stable VdP oscillators have been conducted \cite{defontaines1990chain, shiroky2020nucleation}, demonstrating fundamental regimes of localization and front propagation.

In scientific literature, the phenomenon of synchronization in systems of coupled oscillators is well-documented and exhibits a wide range of variations \cite{rand1980bifurcation, pikovsky2001synchronization, chakraborty1988transition, manevitch2013non, kovaleva2016nonconventional}. The well-known phenomenon of synchronization is characterized by stationary amplitudes and phase shifts, as initially observed by Huygens \cite{huygens1986pendulum}. Commonly, such oscillations are synchronized either in-phase or anti-phase, similar to the Nonlinear Normal Modes (NNMs) of conservative systems \cite{manevitch2011tractable}. Specifically, anti-phase synchronization was also observed by Huygens \cite{huygens1986pendulum}, where the entire system behaves as a collective, single degree-of-freedom (DOF) oscillator. The similarity with NNMs offers a valuable tool for system reduction, which may explain why conventional synchronization has predominantly been the focus of previous studies.
For coupled VdP oscillators, studies \cite{rand1980bifurcation, chakraborty1988transition} and others have identified synchronization regimes where the excitation amplitude of each oscillator remains constant. The behavior of strongly coupled VdP oscillators is examined in \cite{storti1982dynamics}, with comparisons to weakly coupled systems. Recent research classifies this synchronization as being similar to NNM \cite{manevitch2013non, vakakis1996normal}. Recently, more attention has focused on non-stationary regimes involving strong energy exchange in coupled oscillatory systems \cite{manevitch2007new, manevitch2010limiting, manevitch2011tractable}. In more details, recent studies \cite{manevitch2013non, kovaleva2016nonconventional, kovaleva2017non} have uncovered a unique regime of non-stationary "non-conventional" synchronization in systems of linearly coupled bi-stable VdP oscillators. In this regime, the responses of the individual oscillators are strongly modulated, the excitation being exchanged between them. This behavior is fundamentally different from classical synchronization, typically involving the stationary responses. The analytical approach in \cite{manevitch2013non} utilized a special symmetry identified in the system's slow-flow equations for a particular set of parameters. This allowed the slow-flow system to be reduced to a phase plane analysis, where the "non-conventional" synchronization regime was clearly characterized as a limit cycle of the slow flow. Subsequent numerical studies revealed that this non-conventional synchronization regime was robust across a finite range of parameter values, not limited to the specific parameter set that provided the symmetry in the slow-flow equations. The study in \cite{kovaleva2016nonconventional} investigated the impact of detuning in a similar model, while \cite{kovaleva2017non} examined additional effects of damping and non-linearity. The findings indicated that, for the same parameter sets, both stationary synchronous dynamics and "non-conventional" synchronization could coexist.
An alternative approach to modeling two linearly coupled bi-stable VdP oscillators was demonstrated in \cite{shiroky2021modal}. This approach reveals that in the regimes of "non-conventional" synchronization, the amplitudes in the modal coordinates of the system are synchronized, although there is a persistent phase drift. It was also shown that averaging in the modal space substantially simplifies the analysis and removes singularities from the slow-flow equations. Furthermore, these slow-flow equations in the modal space possess sufficient internal symmetry to reduce the dynamics to a simple phase cylinder for any set of control parameters and to demonstrate directly the generic nature of the modal synchronization.

In the aforementioned studies, the self-excitation terms were of the same order of magnitude that the coupling between the oscillators. Current study explores an alternative parametric regime, that, to our opinion, can simplify analysis and understanding of the non-stationary dynamics in the coupled BVdP oscillators. Namely, the coupling is assumed to be small, but the self-excitation is assumed to be even smaller. Thus, in the limit of absent self-excitation, one encounters a multi-parametric degenerate family of stationary beatings and beat waves. The non-zero self-excitation removes the degeneracy and, generically, converts some of the beating tori in the state space into the stable attractors. Numeric and simple analytic exploration reveal that for basic system with few ring-coupled BVdP oscillators these stable tori comprise two eigenmodes (or eigenwaves) with equal amplitudes. Thus, one obtains the natural explanation of the modal synchronization reported in \cite{shiroky2021modal}. However, for a particular case of 4 coupled oscillators one observes a peculiar strong resonance in in the slow-flow state space; this resonance results in more complicated three-wave synchronization patterns.

The structure of the paper is a s follows. Section 2 is devoted to description of the model and detailed analysis of the two-DOF system. Section 3 deals with the generalization to the multi-DOF system. Section 4 treats the special case with internal resonances, and is followed by discussion and concluding remarks in Section 5. 

\section{Model description}
\subsection{Problem formulation}
In this study, a system of two identical, linearly coupled, bi-stable VdP oscillators is investigated, where $m, k,$ and $c$ are the mass, linear stiffness and linear coupling, respectively. $\mu$ is the coefficient of linear and non-linear damping and $\eta$ is the coefficient of the quartic non-linear term.
\begin{align}
\begin{split}
m \Ddot{y_1}  + k y_1 + c (y_1 - y_2) + \mu \Dot{y_1} (1 - y_1 ^2 + \eta y_1 ^4) &= 0\\
m \Ddot{y_2} + k y_2 + c (y_2 - y_1) + \mu \Dot{y_2} (1 - y_2 ^2 + \eta y_2 ^4) &= 0
\end{split}
\end{align} 
We start by a simple scaling of the system, we define the non-dimensional time (t), as $t = \Omega \Tilde{t}$, $ \Omega = \sqrt{\frac{k}{m}}$, and note $'$ as the derivative by the non-dimensional time:
\begin{align}
\begin{split}
&
\begin{cases}
\Omega ^2  y_1'' + \Omega ^2  y_1 +  \frac{c}{m \Omega^2} (y_1 - y_2) +\Omega \frac{\mu}{k} y_1' (1 - y_1 ^2 + \eta y_1 ^4) &= 0\\
\Omega^2 y_2 '' + \Omega^2 y_2 + \frac{c}{m \Omega} (y_2 - y_1) + \Omega \frac{\mu}{k} y_2' (1 - y_2 ^2 + \eta y_2 ^4) &= 0
\end{cases}
\\
&
\begin{cases}
y_1'' + y_1 +  \frac{c}{k} (y_1 - y_2) + \frac{\mu}{\sqrt{k m}} y_1' (1 - y_1 ^2 + \eta y_1 ^4) &= 0\\
y_2 '' + y_2 + \frac{c}{k} (y_2 - y_1) + \frac{\mu}{\sqrt{km}} y_2' (1 - y_2 ^2 + \eta y_2 ^4) &= 0
\end{cases}
\end{split}
\end{align} 
Our model assumptions include weak coupling relative to the linear stiffness ($\frac{c}{k} \ll 1$) and weak non-linearity ($\frac{\mu}{\sqrt{k m}} \ll 1$). In contrast to previous studies \cite{shiroky2021modal}, in this study, we chose to rescale the equations in a way where the control parameter is the coefficient of the nonlinear terms. Therefore, we define $\frac{c}{k} = \varepsilon$ and $\delta = \Omega \frac{\mu}{c}$. Essentially, our choice of scaling is motivated by our interest in studying the effect of the nonlinear terms on the phenomenon, rather than the effect of the coupling. It is also worth noting that one of our key assumptions is that $0 < \varepsilon \ll \delta < 1$. Finally, the non-dimensional equations are obtained:
\begin{align}
\label{2dof_simple_1}
\begin{split}
y_1 ''  + y_1 + \varepsilon (y_1 - y_2) + \varepsilon \delta y_1 '(1 - y_1 ^2 + \eta y_1 ^4) &= 0\\
y_2 '' + y_2 + \varepsilon (y_2 - y_1) + \varepsilon \delta y_2 '(1 - y_2 ^2 + \eta y_2 ^4) &= 0
\end{split}
\end{align} 

\subsection{2 DOF system}
In this section, we will examine the basic 2 DOF system (\ref{2dof_simple_1}) using both numerical simulations and analytical methods. We will begin by numerically investigating the possible motion regimes. The following figure presents these possible regimes:
\begin{figure}[H]
    \centering
    \includegraphics[width=0.9\textwidth]{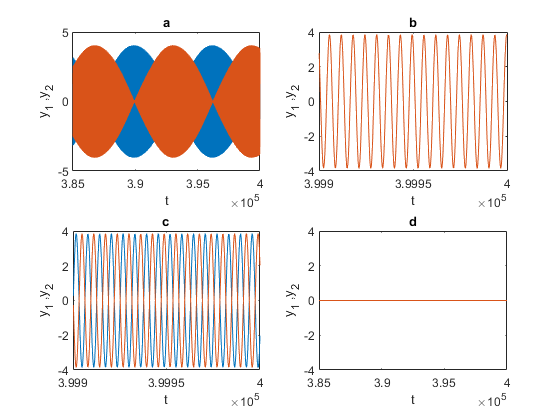}
    \caption{The simulations were performed for parameters $\varepsilon = 0.001, \delta = 0.1 , \eta = 0.1$, with different initial conditions. In figure a we can see beating regimes, figure b is characterized with a rigid body motion, in figure c the system is oscillating in anti-phase, and in figure d the mute mode is presented. }
\end{figure}
The following figure presents the regime distribution based on the initial conditions: 
\begin{figure}[H]
    \centering
    \includegraphics[width=0.9\textwidth]{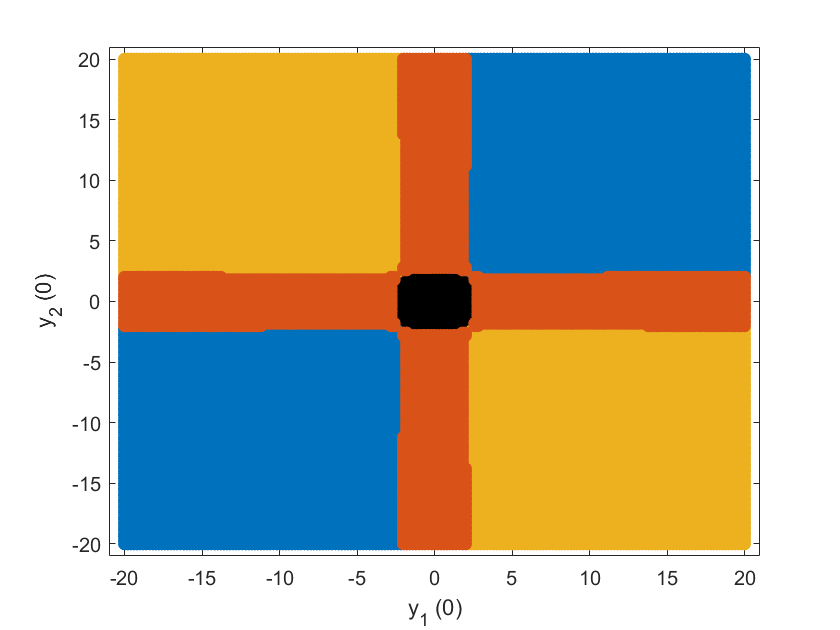}
    \caption{This figure presents the regime distribution for the initial conditions, with  $\dot{y}_1(0) = 0, \dot{y}_2(0) = 0$ and $y_1 (0) , y_2(0)$ as the axis.  Simulations were performed for parameters $\varepsilon = 0.001, \delta = 0.1 , \eta = 0.1$. black, mute mode; yellow, anti-phase motion; blue, rigid body motion; red, beating regime. }
\end{figure}

As shown in \cite{kovaleva2017non}, multiple regimes can coexist in such a system. This study focuses on the beating regime explored in \cite{shiroky2021modal}. Following \cite{shiroky2021modal}, we will perform numerical simulations of the system in both physical and modal coordinates:
\begin{figure}[H]
    \centering
    \includegraphics[width=0.9\textwidth]{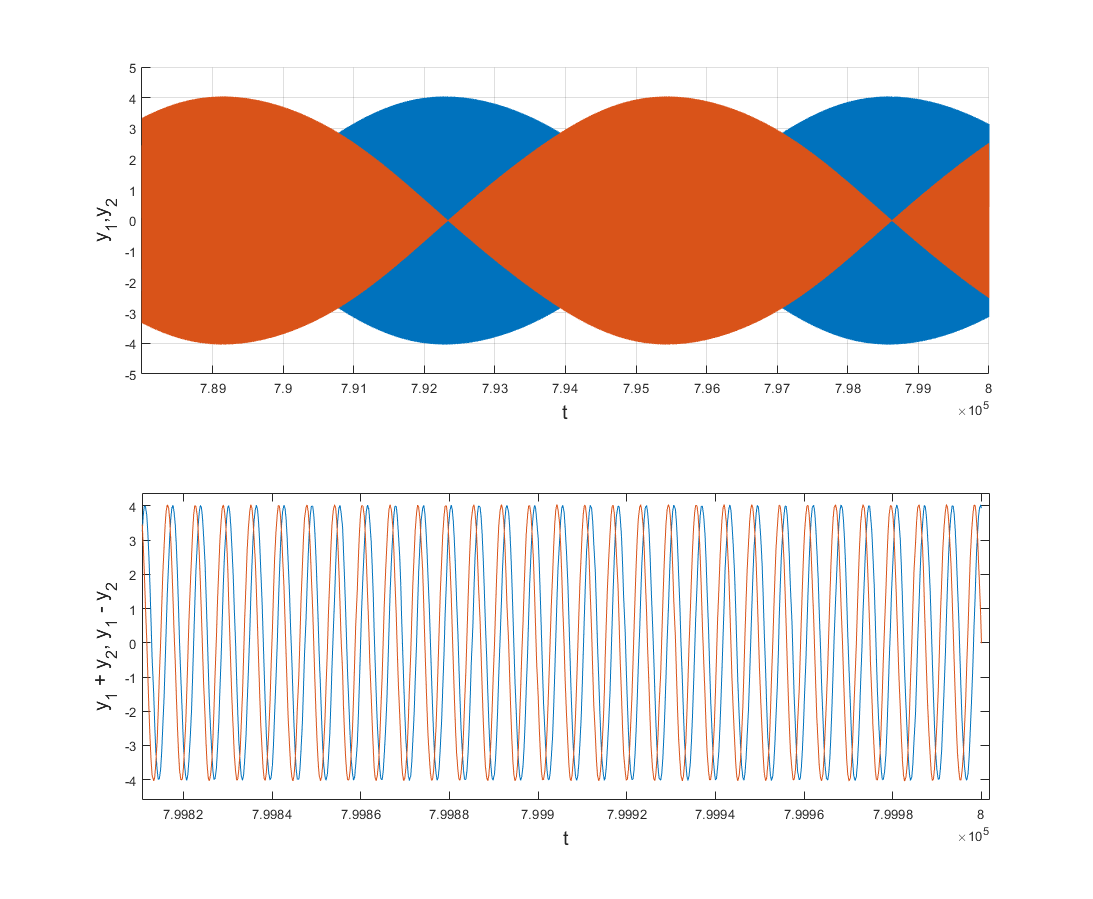}
    \caption{Numerical results from simulations of the system (\ref{2dof_simple_1}), for the parameters $\varepsilon = 0.001, \delta = 0.1 , \eta = 0.1$, for the initial conditions $y_1(0) = 10, \dot{y}_1(0) = 0, y_2(0) = 0, \dot{y}_2(0) = 0 $. The top figure presents the results in the physical coordinates, and the bottom figure presents them in the modal coordinates.}
    \label{2dof_simple_fig_1}
\end{figure}
As observed, a beating pattern emerges in the physical coordinates. Conversely, in the modal coordinates, we observe straightforward results: simple oscillations with a constant amplitude. To preform our analysis, we employ a first-order multiple scales decomposition on the complete system (\ref{2dof_simple_1}), as shown in the relationship below:   
\begin{equation}
\label{multiple_scales_1}
\begin{split}
\varphi_k \exp (it) = \dot{y_k} + i y_k , k=1,2\\ 
\end{split}
\end{equation}
A slower time scale is defined, $\tau = \varepsilon t$. The following secular equations describe the slow beating envelope:
\begin{equation}
\label{modulations_2dof_simple_1}
\begin{split}
\varphi_1 ' - \frac{i}{2} (\varphi_1 - \varphi_2) + \frac{\delta \varphi_1}{2} \left(1 - \frac{|\varphi_1 ^2|}{4} + \frac{|\eta \varphi_1 ^4|}{8} \right) = 0  \\ 
\varphi_2 ' - \frac{i}{2} (\varphi_2 - \varphi_1) + \frac{\delta \varphi_2}{2} \left(1 - \frac{|\varphi_2 ^2|}{4} + \frac{|\eta \varphi_2 ^4|}{8} \right) = 0
\end{split}
\end{equation}
As shown in Figure (\ref{2dof_simple_fig_1}), the amplitude in the modal plane is constant. Therefore, we will analyze the system using modal coordinates. The following modal transformation and its inverse \cite{shiroky2021modal} will be employed:
\begin{equation}
\begin{split}
&\mu_0 = \varphi_1 + \varphi_2 \; \; , \mu_1 = \varphi_1 - \varphi_2 \\
&\varphi_1 = \frac{\mu_0 + \mu_1}{2} , \varphi_2 = \frac{\mu_0 - \mu_1}{2}
\end{split}
\end{equation}
It should also be noted that each modal coordinate evolves according to the following relation as \(\delta \to 0\), where \(\Omega_j\) is the linear natural frequency of the slow-varying system (\ref{modulations_2dof_simple_1}):
\begin{equation}
\begin{split}
&\mu_0' = \Omega_0 \mu_0,\; \; \mu_1' = \Omega_1 \mu_1,\; \; \Omega_0 = 0, \; \;  \Omega_1 = 1
\end{split}
\end{equation}
After substitution into the modulations equations (\ref{modulations_2dof_simple_1}) and few algebraic manipulations we obtain the equations in the modal coordinates:
\begin{equation}
\label{modal_equaiton_2dof_1}
\begin{split}
&\mu_0 ' = -\frac{\delta}{256} \Bigg(  \eta  \mu _0^3 \left( \mu _0 ^{* ^2} + \mu _1 ^{*^2}\right)+2  \mu _0^2 \mu _0^ * \left(3 \eta  \mu _1 \mu _1{}^* -4\right) + \\  
&\mu _0 \left(3 \eta  \mu _1^2 \mu_0 ^{* ^2}+3 \eta  \mu _1^2 \mu _1 ^* {}^2 - 16 \mu _1 \mu _1 {}^* + 128\right)  +2 \mu _1^2 \mu _0{}^* \left(\eta  \mu _1 \mu _1 ^{*} -4\right) \Bigg)
\\
&\mu_1' = i \mu_1 - \frac{\delta}{256} \Bigg(\eta  \mu _1^3 \mu_0 ^{* ^2} +\mu _1 \left(3  \eta  \mu _0^2 \mu_0^{*^2} - 16  \mu _0 \mu _0 ^* +128 \right)+ \\ 
&\eta  \left(3 \mu _0^2+\mu _1^2\right) \mu _1 \mu _1^{*^2}+2 \mu _1{}^* \left(\eta  \mu _0^3 (\mu _0{}^*+3 \eta  \mu _1^2 \mu _0 \mu _0 {}^* -4 \mu _0^2-4 \mu _1^2\right)\Bigg)
\end{split}
\end{equation}
One can identify two time scales, the obvious one which is the slow time, $\tau = \varepsilon t$, and the super slow time scale $\xi = \delta \tau$. Ansatz both variables will take the following form: 
\begin{equation}
\label{an_2dof_1}
\begin{split}
\mu_j = \psi_j(\xi) \exp(i \Omega_j \tau)
\end{split}
\end{equation}
Substituting (\ref{an_2dof_1}) into the modal equations (\ref{modal_equaiton_2dof_1}), a set equations is obtained:
\begin{align}
\label{exp_2dof_1}
\begin{split}
& \frac{d \psi_0}{d \xi} = -\frac{1}{256} \Bigg(\eta  \psi_0 \psi_0^{*^2} \left({\psi_0}^2+3 \exp({2 i \tau}) {\psi _1}^2\right)+2 \psi_0^* \left(\eta  \psi_1 \psi_1^* \left(3 \psi_0 ^2 \right. \right. \\ 
&\ \left. \left.+\exp({2 i \tau}) \psi_1 ^2\right) -4 \left(\psi_0 ^2+\exp({2 i \tau}) \psi_1 ^2\right)\right)\\
&+ \eta  \psi_0 \psi_1 ^{* ^2} \left(3 \psi_1^2+\exp({-2 i \tau}) \psi_0 ^2 \right) -16 \psi_0 \psi_1 \psi_1^* + 128 \psi_0 \Bigg) \\
&\frac{d \psi_1}{d \xi} = 
- \frac{1}{256} \Bigg(\eta \psi_1  \psi_0 ^{* ^2} \left(3 \psi_0 ^2+\exp({2 i \tau}) \psi_1^2 \right)
+2 \exp({-2 i \tau}) \psi_0 \psi_0^* \left(\eta  \psi_1 ^* \left(\psi_0 ^2
\right. \right. \\ & \left. \left.+ 3 \exp({2 i \tau}) \psi_1 ^2\right) -8 \psi_1 \right) +\eta \exp({-2 i \tau}) \psi_1 \psi_1^{*^2} \left(3 \psi_0 ^2+\exp({2 i \tau}) \psi_1^2\right) \\ 
&-8 \exp({-2 i \tau}) \psi_1^* \left(\psi_0 ^2+\exp({2 i \tau}) \psi_1 ^2\right)
+128 \psi_1 \Bigg)
\end{split}
\end{align}
As one can see, some of the terms are multiplied by rapidly oscillating terms (according to the super-slow time scale) . In order to simplify the problem, we define the averaged slow-flow envelopes as $\Psi_j = \left< \psi_j \right>$. The equations for the slow flow envelope are obtained by taking equations (\ref{exp_2dof_1}) and removing the rapidly oscillating terms:
\begin{align}
\label{avg_2dof_1}
\begin{split}
&\frac{d \Psi_0}{d \xi} = -\frac{1}{256} \left( \eta  \Psi_0 ^3 \Psi_0 ^{*^2} +2 \Psi_0 ^2 \Psi_0 ^* \left(3 \eta  \Psi_1 \Psi_1^* -4\right) +3 \eta  \Psi_0 \Psi_1 ^2 \Psi_1 ^{* ^2} \right. 
\\
& \left. -16 \Psi_0 \Psi_1 \Psi_1 ^* +128\Psi_0 \right) \\
&\frac{d \Psi_1}{d \xi} = 
- \frac{1}{256} \left(\eta   \Psi_1 ^3  \Psi_1 ^{* ^2} +2  \Psi_0  \Psi_0 ^* \Psi_1  \left(3 \eta   \Psi_1  \Psi_1^*-8 \right)+3 \eta  \Psi_0^2 \Psi_0 ^{* ^2} \Psi_1 \right. 
\\
&\left. -8  \Psi_1^2 \Psi_1^* +128 \Psi_1 \right)
\end{split}
\end{align}
We look for a stationary solution in the modal space, therefore the we set the derivatives by the time to zero, and decompose the the solution into modulus and argument parts:
\begin{equation}
\label{comp_de_2dof_1}
\begin{split}
&\Psi_0 = \Psi_1 = \sqrt{Z} \exp(i\Delta) \\
&\frac{\Psi_0}{d \xi} = \frac{\Psi_1}{d \xi} = 0
\end{split}
\end{equation}
Substituting (\ref{comp_de_2dof_1}) into the (\ref{avg_2dof_1}), two identical algebraic equations are obtained:
\begin{align}
\begin{split}
\label{2_dof_solution}
&64 - 12 Z ^2 + 5 Z ^4 \eta = 0 \\
&Z = 4 \; \frac{3 \pm \sqrt{9 - 80 \eta} }{10 \eta} 
\end{split}
\end{align}
Therefore, the modal amplitude will be as follows: $A = 2 \sqrt{\frac{3 \pm \sqrt{9 - 80 \eta} }{10 \eta}}$, although according to numerical simulations only the upper solution is stable. In order to find the beating envelope, we can return to the physical coordinates using the inverse transformation. We know the following relationship exists:
\begin{align}
\begin{split}
&\varphi_0  \approx \frac{\Psi_0 + \Psi_1 \exp(i \tau)}{2} = \frac{A + A \exp(i \tau)}{2} \\
&\varphi_1 \approx \frac{\Psi_0 - \Psi_1 \exp(i \tau)}{2} = \frac{A - A \exp(i \tau)}{2}
\end{split}
\end{align}
Thus:
\begin{align}
\label{approx_env_2dof_1}
\begin{split}
&|\varphi_0|  \approx \sqrt{ \frac{A^2}{2} + \frac{A^2}{2} cos(\tau)}  \\
&|\varphi_1|  \approx \sqrt{\frac{A^2}{2} - \frac{A^2}{2} cos(\tau)}
\end{split}
\end{align}
In the following figure, a numerical verification of the results is presented:
\begin{figure}[H]
    \centering
    \includegraphics[width=0.9\textwidth]{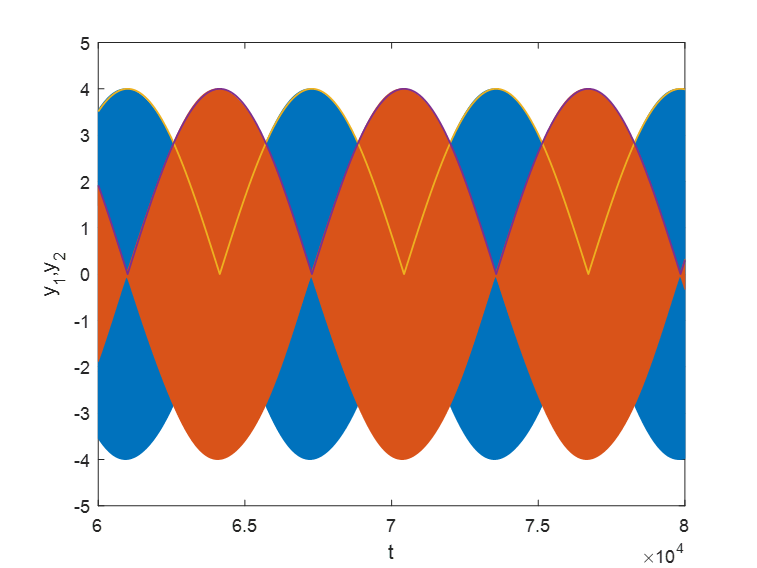}
    \caption{Numerical results from simulations of the system (\ref{2dof_simple_1}), for the parameters $\varepsilon = 0.001, \delta = 0.1 , \eta = 0.1$, for the initial conditions $y_1(0) = 10, \dot{y}_1(0) = 0, y_2(0) = 0, \dot{y}_2(0) = 0 $. (blue,red) - displacements of particles 1,2 respectively. (yellow,purple) - the analytical approximation (\ref{approx_env_2dof_1}) for particles 1,2 respectively.}
\end{figure}
\newpage
\section{Multiple DOF} \label{avg_analysis_1_sec}
In this section a chain of N weakly coupled identical bi-stable VdP oscillators will be investigated:
\begin{align}
\label{chain_simple_1}
\begin{split}
&y_n'' + y_n + \varepsilon (2y_n - y_{n-1} - y_{n+1}) + \varepsilon \delta y_n '(1 - y_n ^2 + \eta y_n ^4) = 0, \;\;  y_{n} = y_{n+N}\\
\end{split}
\end{align}
The solution for the free system (without the nonlinear terms) is of the following form:
\begin{align}
\begin{split}
\label{k_def_1}
&y_{n} = y_{n+N}, \; \; \; k_l N = 2\pi l,\; \; \; l \in  \mathbf{Z} \; \; \; \; \; \; \; \; \; \; \; \; \; \; \\
 &k_l = \frac{2\pi l}{N}, l = 0, 1, ... , N-1 ;\; \; \; \omega^2 = 1 + 4 \varepsilon c \sin^2\left(\frac{k_l}{2}\right)
\\
\end{split}
\end{align}
\\
We used the free system's solution to set the initial conditions as a linear combination of waves for $l = 0, l = 1$ with zero amplitudes and zero initial phase shift:
\begin{align}
\begin{split}
\label{in_cond_1}
&y_0 = A \begin{pmatrix} \cos(k_0 - \omega_0 t)+\cos(k_1 - \omega_1 t) \\ \vdots \\ \cos(k_0 n - \omega_0 t)+\cos(k_1 n - \omega_1 t) \end{pmatrix} \; ;N=3 \\
&\Dot{y}_0 = A \begin{pmatrix} \omega_0 \sin(k_0 - \omega_0 t)+\omega_1 \sin(k_1 - \omega_1 t) \\ \vdots \\ \omega_0 \sin(k_0 n - \omega_0 t)+\omega_1 \sin(k_1 n - \omega_1 t) \end{pmatrix} \; ;N=3
\end{split}
\end{align}
For simpler analysis, the case of $N=3$ will be examined. For this case, the following solution is obtained:
\begin{align}
\begin{split}
\label{sol3dof_1}
y_n(t) = 2A\left( \cos\left(\frac{(\sqrt{1+3 \varepsilon} -1)t }{2} -\frac{\pi n}{3}\right) \cos\left(\frac{(\sqrt{1+3 \varepsilon} + 1)t }{2} -\frac{\pi n}{3}\right) \right)
\end{split}
\end{align}
The beating envelope is determined by the slow modulation in (\ref{sol3dof_1}), therefore:
\begin{equation}
\begin{split}
\label{env_3dof_1}
Amp(y_n(t)) = 2A\left| \cos\left(\frac{(\sqrt{1+3 \varepsilon} -1)t }{2} -\frac{\pi n}{3}\right) \right|; n = 1,2,3
\end{split}
\end{equation}
The results of the numerical simulations are presented in the following figure:
\begin{figure}[H]
    \centering
    \includegraphics[width=0.9\textwidth]{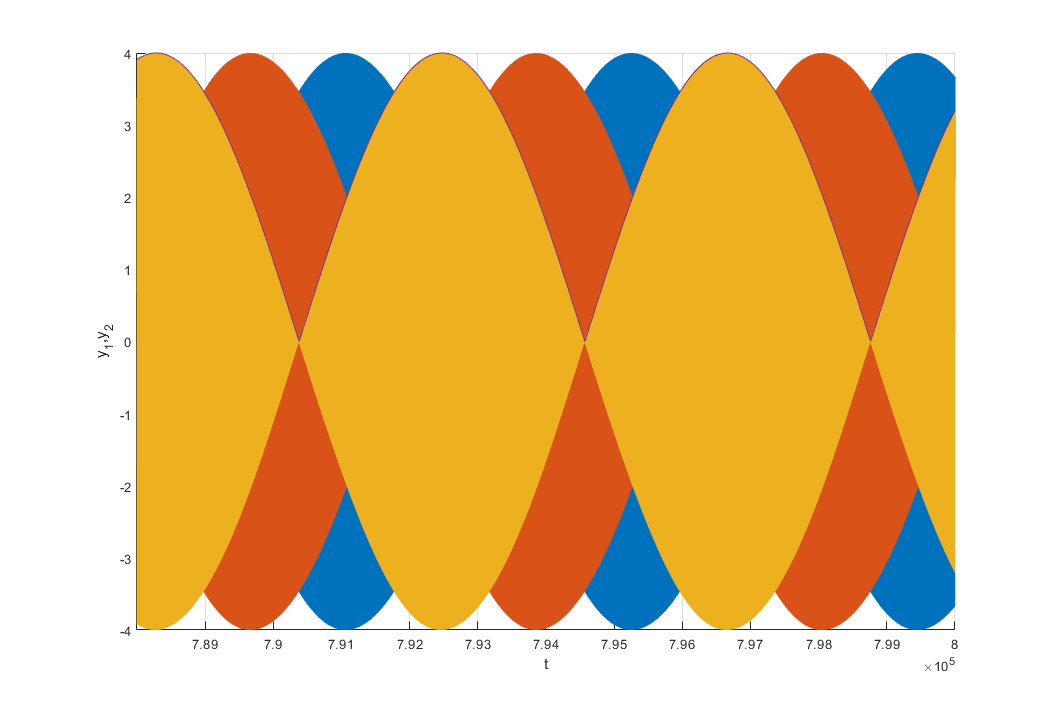}
    \caption{Time series for N=3, initial conditions as in (\ref{in_cond_1}), (blue,red,yellow) - displacements of particles 1,2,3 respectively, $\varepsilon = 0.001, \delta = 0.1 , \eta = 0.1$. The purple envelope is evaluated according to (\ref{env_3dof_1}),for n=3 with insignificant phase shift for A=2.}
    \label{3dof_fig_full_1}
\end{figure}
The movement of the wave traveling through the ring clearly shows a pattern consistent with the simple superposition of eigenwaves for \(l = 0\) and \(l = 1\) with equal amplitudes. This implies that the case \(N = 3\) is a straightforward generalization of the case \(N = 2\), one simply uses eigenwaves instead of eigenmodes. The value of the envelope amplitude \(A = 2\) mentioned in the caption of Figure (\ref{3dof_fig_full_1}) will be explained later. Averaging, or first-order multiple scales decomposition, is performed on the system (\ref{chain_simple_1}) in accordance with the following relationship, as done in the simpler cases:
\begin{equation}
\begin{split}
\varphi_k \exp({it}) = \Dot{y} _k + i y_k , k = 1,...N
\end{split}
\end{equation}
In the basic order of approximation, the secular equations that  describe the slow modulation envelopes in system (\ref{chain_simple_1}) take the following form:
\begin{align}
\begin{split}
\label{avg_3dof_1}
&\varphi_k ' -\frac{i}{2} (2 \varphi_k - \varphi_{k+1} - \varphi_{k-1}) +\delta f_k = 0, k = 1,...N\\
&\varphi_0 \equiv \varphi_N, \varphi_N+1 \equiv \varphi_1, f_k = \frac{\varphi_k}{2} \left(1 - \frac{|\varphi_k|^2}{4} + \eta \frac{|\varphi_k|^4}{8}\right) 
\end{split}
\end{align}
The apostrophe denotes differentiation with respect to the slow time scale, $\tau = \varepsilon t$. We introduce the relationship between the original coordinates to the absolute value of the slow-varying system's coordinates:
\begin{align}
\begin{split}
 \Dot{y} _k + i y_k = \left|\varphi_k \right| \exp({i(t +arg(\varphi_k))}) , k = 1,...N
\end{split}
\end{align}
This relationship shows that the modulation envelope is described by \(\left| \varphi_k \right|\), as illustrated in the following figure:
\begin{figure}[H]
    \centering
    \includegraphics[width=0.9\textwidth]{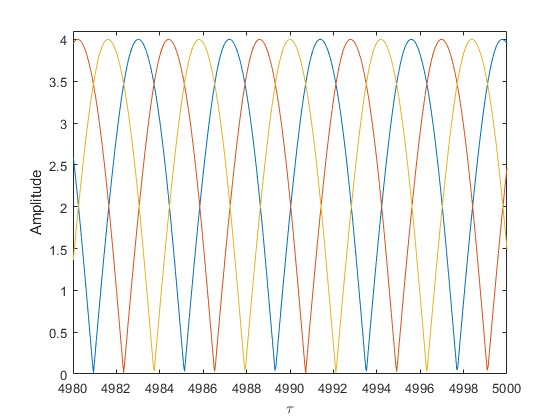}
    \caption{Modulation envelope $\left| \varphi_k \right|$ versus the slow time scale $\tau$ for $N = 3, \delta = 0.1, \eta = 0.1$. The envelopes for the particles are denoted by colors (blue, red ,yellow) respectively. Initial conditions: $\varphi_1 (0) = 4 , \varphi_2 (0) = 3 -i\sqrt{3}, \varphi_3 (0) = 2-\sqrt{3} + i$.}
    \label{3dof_fig_env_1}
\end{figure}
For further simplification and a better understanding of the phenomenon, we present the results in modal coordinates. The modal coordinates $\mu_j$ are defined based on system (\ref{avg_3dof_1}) in the limit as $\delta \to 0$:
\begin{equation}
\begin{split}
\label{3mod_def_1}
\varphi_k = \sum_{j=0} ^{N-1} \mu_j \exp \left({\frac{2\pi i (k-1) j}{N}}\right); \mu_j = \sum_{k=1} ^{N} \varphi_k \exp \left({\frac{ -2\pi i k j}{N}}\right)  ; j = 0,...N-1
\end{split}
\end{equation}
The modal coordinates are defined according to (\ref{3mod_def_1}), and automatically satisfy the periodic boundary condition. In the limit $\delta \to 0$, the modal coordinates evolve according to the following equation:
\begin{equation}
\begin{split}
\mu_j ' = i\Omega_j \mu_j, \Omega_j = 2\sin^2\left(\frac{\pi j}{N}\right)
\end{split}
\end{equation}
Therefore, in the absence of the nonlinear terms ($f_k$ in (\ref{avg_3dof_1}) ) the modal amplitudes oscillate with constant amplitude. When $\delta > 0$ the situation will change as shown in the following figure:
\begin{figure}[H]
    \centering
    \includegraphics[width=0.9\textwidth]{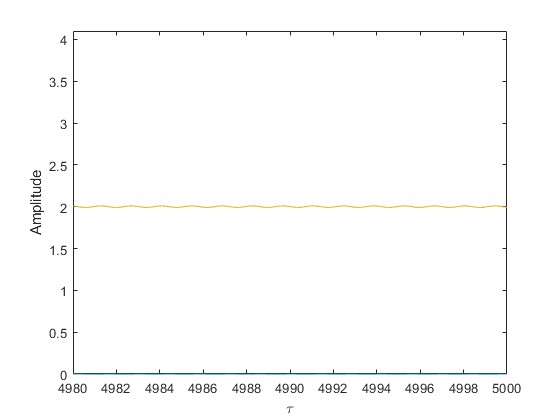}
    \caption{Modal amplitudes for $\mu_j$ for system (\ref{avg_3dof_1}) according to (\ref{3mod_def_1}), $N=3,\delta=0.1,\eta=0.1$, the modal amplitudes are plotted in (blue,red,yellow) respectively. The initial conditions are the same as in (\ref{3dof_fig_env_1}). }
    \label{3dof_mod_amp_1}
\end{figure}
As can be seen in Figure (\ref{3dof_mod_amp_1}), the modal amplitude is almost constant. Thus, the role of the BVdP terms (\(f_k\) in (\ref{avg_3dof_1})) is clear: they first determine the mean values of the modal amplitudes and second, cause minor oscillations around these average constant values, which can be considered small perturbations. To understand the structure of the equations and their solutions, we present them in modal coordinates according to the direct modal transformation (\ref{3mod_def_1}). The derivatives are computed from equation (\ref{avg_3dof_1}) and the inverse modal transformation as described in (\ref{3mod_def_1}). In the next part, the same analysis performed for the simpler 2 DOF case will be conducted. After substituting the derivatives and the modal transformation into the slow modulation equations, we obtain an equation in modal coordinates. Unfortunately, this system is still too complex for a complete analysis. However, two time scales can be identified: the basic time scale \(\tau\) and a super-slow time scale \(\xi = \delta \tau\). The Ansatz for all variables will take the following form:
\begin{equation}
\label{rescale_eq_1}
\mu_j = \psi_j(\xi) \exp\left(i\Omega_j \frac{\xi}{\delta}\right), \mu_j ' = \left(\frac{d \psi_j}{d \xi} +\frac{\Omega_j i \psi _j}{\delta} \right) \delta\exp\left(i\Omega_j \frac{\xi}{\delta} \right) 
\end{equation}
We substitute (\ref{rescale_eq_1}) into the equations in modal coordinates, and the resulting equations become even more complicated. However, it can be observed that some terms in these equations are multiplied by rapidly oscillating terms (with respect to the time $\frac{\xi}{\delta}$). To deal with these factors, we can average them out. We define the averaged slow-flow envelopes as follows:
\begin{equation}
\label{avg_coordinate}
\Psi_j = \left< \psi_j \right>_\frac{\xi}{\delta}
\end{equation}
The equations for the averaged slow-flow envelopes are obtained by removing all terms that oscillate rapidly (i.e., those containing exponents of the time scale \(\frac{\xi}{\delta}\)):
\newpage
\begin{align}
\begin{split}
\label{avg_3dof_eq}
&\frac{d \Psi_0}{d \xi} = -\frac{1}{16} \Bigg( 2 \left(4 \Psi_0 - \Psi_0 ^2 \Psi_{0} ^* - 2\Psi_{0} (\Psi_1 \Psi_1 ^{*} +\Psi_2 \Psi_2 ^{*} )\right)
+\eta (6 \Psi_0 ^2 \Psi_0 ^{*} \left(\Psi_1 \Psi_1 ^* 
\right. \\& 
\left. +\Psi_2 \Psi_2 ^* \right) + \Psi_0^3 \Psi_0^2  +12\Psi_0 \Psi_1 \Psi_1^* \Psi_2 \Psi_2^* + 3\Psi_0 (\Psi_1 ^2 \Psi_1 ^{* ^2} +\Psi_2 ^2  \Psi_2 ^{* ^2}))\Bigg)\\
&\frac{d \Psi_1}{d \xi} = -\frac{1}{16} \Bigg( 2 \left(4 \Psi_1 - \Psi_1 ^2 \Psi_1 ^* - 2\Psi_1 (\Psi_0 \Psi_0 ^* +\Psi_2 \Psi_2 ^* )\right)
+\eta ( \Psi_1 ^ {* ^2} \Psi_2 ^{3} + \Psi_1 ^{3} \Psi_1 ^{* ^2} \\
&+ 6 \Psi_1 ^{2} \Psi_1 ^{*} (\Psi_0 \Psi_0 ^* +\Psi_2 \Psi_2 ^*) +3 \Psi_{1} (\Psi_0 ^2 \Psi_0 ^{* ^2} +\Psi_2 ^2 \Psi_2 ^{* ^2} )+12\Psi_1 \Psi_0 \Psi_0 ^{*} \Psi_2 \Psi_2 ^{*}) \Bigg)
\\
&\frac{d \Psi_2}{d \xi} = -\frac{1}{16} \Bigg( 2 \left(4 \Psi_2 - \Psi_2 ^2 \Psi_2 ^* - 2\Psi_2 (\Psi_0 \Psi_0 ^* +\Psi_1 \Psi_1 ^* )\right)
+\eta ( \Psi_2 ^{*^2} \Psi_1^3 + \Psi_2 ^3 \Psi_2 ^{* ^2} \\
&+ 6 \Psi_2^2 \Psi_2 ^{*} (\Psi_0 \Psi_0 ^* +\Psi_1 \Psi_1 ^*) +3 \Psi_2 (\Psi_0 ^2 \Psi_0 ^{* ^2} +\Psi_1 ^2 \Psi_1 ^{* ^2} )+12\Psi_1 \Psi_0 \Psi_0^* \Psi_2 \Psi_2^*) \Bigg)
\end{split}
\end{align}
The obtained set of equations is simple enough to complete the analysis. In the case of $N=3$, we search for a solution where $\Psi_1 \equiv 0$. The system is reduced to the following form:
\begin{align}
\begin{split}
&\frac{d \Psi_0}{d \xi} = -\frac{1}{16} \left( 2 \left(4 \Psi_0 - \Psi_0 ^2 \Psi_0 ^* - 2\Psi_0 ( \Psi_2 \Psi_2 ^* )\right)
+\eta (6 \Psi_0^2 \Psi_0^* (\Psi_2 \Psi_2 ^*) + \Psi_0^3 \Psi_0^2 \right.
 \\
 & \left. + 3\Psi_0 (\Psi_2 ^2  \Psi_2 ^{* ^2}))\right)
\\
&\frac{d \Psi_1}{d \xi} = 0
\\
&\frac{d \Psi_2}{d \xi} = -\frac{1}{16} \left( 2 \left(4 \Psi_2 - \Psi_2 ^2 \Psi_2 ^* - 2\Psi_2 (\Psi_0 \Psi_0 ^* )\right)
+\eta ( \Psi_2 ^3 \Psi_2 ^{* ^2} + 6 \Psi_2^2 \Psi_2^* (\Psi_0 \Psi_0 ^* ) \right)
\\
&\left. +3 \Psi_2 (\Psi_0 ^2 \Psi_0 ^{* ^2} )) \right)
\end{split}
\end{align}
As one can see, the system becomes completely symmetric with respect to ($\Psi_0, \Psi_2$). We look for a stationary solution in the modal space, and decompose it into modulus and argument parts:
\begin{equation}
\begin{split}
\Psi_0 = \Psi_2 = \sqrt{Z} \exp(i\Delta) = const
\end{split}
\end{equation}
After simple calculations, the obtained solution is similar to the solution of the 2 DOF system (\ref{2_dof_solution}). For the case considered in the simulations, where $\eta=0.1$, and consequently $A=\sqrt{Z} = 2$ the analytical approximation matches the numerical simulations shown in Figure (\ref{3dof_fig_full_1}). Note that only the solution with the upper sign yields a stable result.

\section{Internal resonances}
In the case of $N=4$, a different regime is obtained, as demonstrated in the following figures:
\begin{figure}[H]
    \centering
    \includegraphics[width=0.7\textwidth]{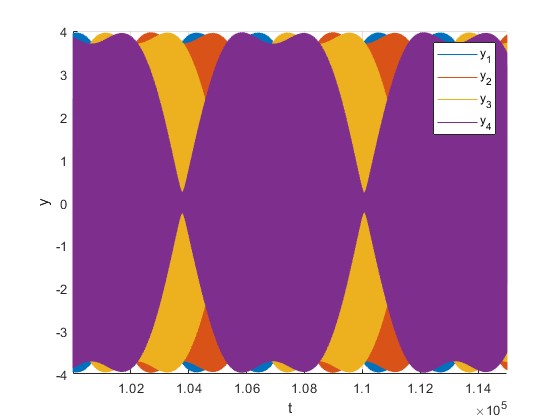}
    \caption{The regime that was obtained from simulating the system for (\ref{in_cond_1}) as the initial conditions, with the parameters $\eta = 0.1, \delta = 0.1, \varepsilon = 0.001$} 
    \includegraphics[width=0.7\textwidth]{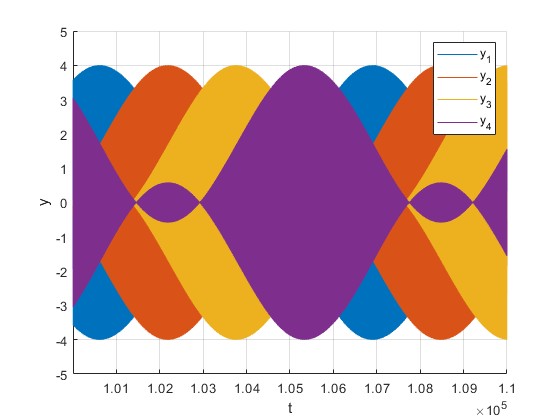}
    \caption{The regime that was obtained from simulating the system for $y_0 = \begin{pmatrix} 0.531067 \\ 1.735244 \\ 0.079833 \\ -1.607738 \end{pmatrix}, \Dot{y_0} = \begin{pmatrix} -3.936128\\  0.234127 \\ -0.5917038\\ -0.216912\end{pmatrix} $ as the initial conditions, with the parameters $\eta = 0.1, \delta = 0.1, \varepsilon = 0.001$}
    \label{4_dof_dbl}
\end{figure}
We can use the same formulation that we used for the general case to approximate the modulation envelope, with some modifications to fit the model to the observed phenomenon. Firstly, we will examine a simulation of the averaged system in both physical and modal coordinates, similar to Figure (\ref{3dof_mod_amp_1}):
\begin{figure}[H]
    \centering
    \includegraphics[width=1\textwidth]{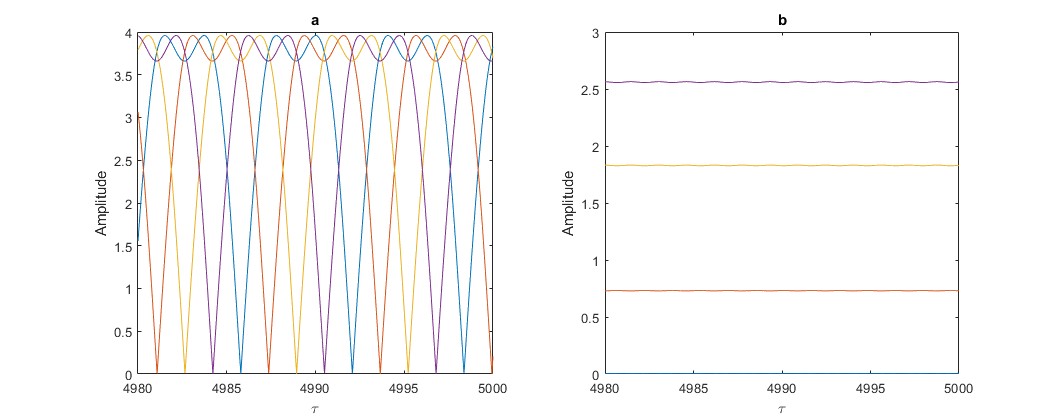}
    \caption{Modulation envelope of $\left| \varphi_k \right|$ in figure a, and the modal amplitude of $\left| \mu_j \right|$ in figure b  versus the slow time $\tau$ for $N=4, \delta = 0.1, \eta = 0.1$, particles 1,2,3,4 and the modal amplitude of modes 1,2,3,4 are denoted by blue, red, yellow and purple respectfully. The initial conditions:
    $\varphi_1 (0) = 0.049011+0.655865i,\varphi_2 (0) = 3.480829+1.756901i,\varphi_3 (0) = 3.677987+0.178928i,\varphi_4 (0) = 3.004952-1.872470i  $} 
    \label{modal4a_1}
    \includegraphics[width=1\textwidth]{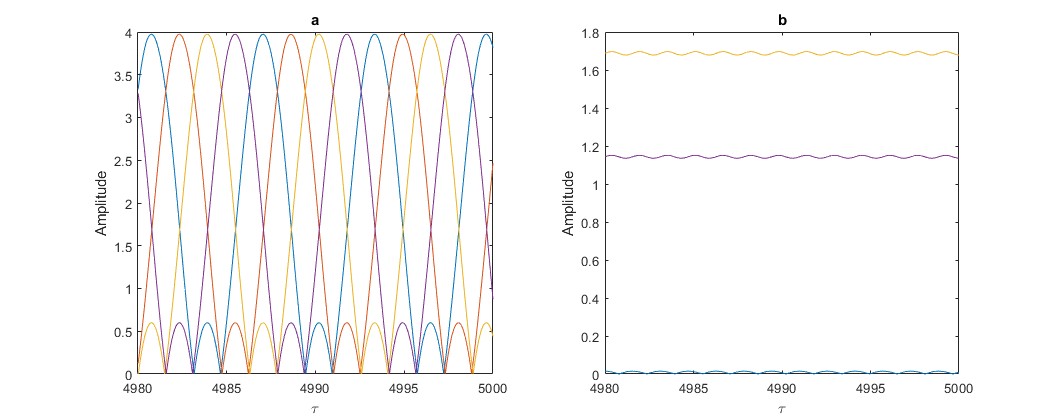}
    \caption{Modulation envelope of $\left| \varphi_k \right|$ in figure a, and the modal amplitude of $\left| \mu_j \right|$ in figure b versus the slow time $\tau$ for $N=4, \delta = 0.1, \eta = 0.1$, particles 1,2,3,4 and the modal amplitude of modes 1,2,3,4 are denoted by blue, red, yellow and purple respectfully. The initial conditions:
    $\varphi_1 (0) = 4,\varphi_2 (0) = 2-2i,\varphi_3 (0) = 0,\varphi_4 (0) = 2+2i  $}
    \label{modal4b_1}
\end{figure}
As shown in figures (\ref{modal4a_1},\ref{modal4b_1}),the modal amplitudes  are almost constant. This demonstrates the role of the BVdP terms as discussed previously. We perform the same analysis that was done for the case of N=3. In this case, the equations are far more complicated then those analysed for N=3. It's important to note that for N=4, one encounters $\Omega_0 = 0, \Omega_1 = \Omega_3 = 1, \Omega = 2$. Consequently, multiple options for the resonant  terms arise. After obtaining the slow-modulations equations in modal coordinates, we substitute  (\ref{rescale_eq_1}) into the modal equations and average out the rapidly oscillating terms using (\ref{avg_coordinate}).
As observed in figures (\ref{modal4a_1},\ref{modal4b_1}), $\Psi_1 \equiv 0$. Therefore, we look for a solution that upholds $\Psi_1 \equiv 0$ and obtain a bit simpler set of equations:
\begin{equation}
\begin{split}
\label{final_eq_4dof}
&\frac{d\Psi_0}{d\xi} = -\frac{1}{16} \Bigg( 2 \Bigg[ 4\Psi_0 - \Psi_0 ^2 \Psi_0 ^* - 2\Psi_0 \left( \Psi_2 \Psi_2 ^* + \Psi_3 \Psi_3 ^* \right) - \underline{\Psi_3 ^2  \Psi_2 ^*} \Bigg] \\
&+ \eta \Bigg[ \Psi_0 \Bigg( 3 \left( \Psi_2 ^2  \Psi_2 ^{*^2} + \Psi_3 ^2  \Psi_3 ^{*^2} \right) + 12 \Psi_2 \Psi_2 ^* \Psi_3 \Psi_3 ^*  + 6 \Psi_0 \Psi_0 ^*  \left( \Psi_2 \Psi_2 ^*  + \Psi_3 \Psi_3 ^* \right) \\
& + \Psi_0 ^2 \Psi_0 ^{*^2} \Bigg)
+ \underline{\Psi_2 ^*  \left( 6 \Psi_0 \Psi_0 ^* \Psi_3 ^2  + 3 \Psi_2 \Psi_2^* \Psi_3 ^2 + 2 \Psi_3 ^3 \Psi_3 ^*  \right) + 3\Psi_0 ^2 \Psi_2 \Psi_3 ^*} \Bigg] \Bigg) \\
&\frac{d\Psi_1}{d\xi} = 0 \\
&\frac{d\Psi_2}{d\xi} = -\frac{1}{16} \Bigg( 2 \Bigg[ 4\Psi_2 - \Psi_2 ^2 \Psi_2 ^* - 2\Psi_2 \left( \Psi_0 \Psi_0 ^* + \Psi_3 \Psi_3 ^* \right) - \underline{\Psi_0 ^*  \Psi_3 ^2} \Bigg] \\
&+ \eta \Bigg[ \Psi_2 \Bigg( 3 \left( \Psi_0 ^2  \Psi_0 ^{*^2} + \Psi_3 ^2  \Psi_3 ^{*^2} \right) + 12 \Psi_0 \Psi_0 ^* \Psi_3 \Psi_3 ^*  + 6 \Psi_2 \Psi_2 ^*  \left( \Psi_0 \Psi_0 ^*  + \Psi_3 \Psi_3 ^* \right) \\
& + \Psi_2 ^2 \Psi_2 ^{*^2} \Bigg) + \underline{\Psi_0 ^*  \left( 3 \Psi_0  \Psi_0 ^* \Psi_3 ^2   + 6 \Psi_2 \Psi_2 ^* \Psi_3 ^2 + 2 \Psi_3 ^3  \Psi_3 ^* \right) + 3\Psi_0 \Psi_2 ^2  \Psi_3 ^{*^2}} \Bigg] \Bigg)\\
&\frac{d\Psi_3}{d\xi} = -\frac{1}{16} \Bigg( 2 \Bigg[ 4\Psi_3 - \Psi_3 ^2 \Psi_3 ^* - 2\Psi_3 \left( \Psi_0 \Psi_0 ^* + \Psi_2 \Psi_2 ^* \right) -\underline{ 2 \Psi_0 \Psi_2 \Psi_3 ^*} \Bigg] \\
&+ \eta \Bigg[ \Psi_3 \Bigg( 3 \left( \Psi_0 ^2  \Psi_0 ^{*^2} + \Psi_2 ^2  \Psi_2 ^{*^2} \right) + 12 \Psi_0 \Psi_0 ^* \Psi_2 \Psi_2 ^*  + 6 \Psi_3 \Psi_3 ^*  \left( \Psi_0 \Psi_0 ^*  + \Psi_2 \Psi_2 ^* \right) \\
& + \Psi_3 ^2 \Psi_3 ^{*^2} \Bigg) + \underline{\Psi_3 ^*  \left( 6 \Psi_0 ^2 \Psi_0 ^* \Psi_2 + 6 \Psi_2 ^2  \Psi_2^* \Psi_0 + 6 \Psi_0 \Psi_2 \Psi_3 \Psi_3 ^*  \right) + 2\Psi_0 ^*  \Psi_2 ^* \Psi_3 ^3} \Bigg] \Bigg)
\end{split}
\end{equation}
Note that almost all the terms in equation (\ref{final_eq_4dof}) have a similar form, except for the "resonant" terms, which are underlined. The significance of these terms will be understood in the next part of the analysis. As in (\ref{avg_analysis_1_sec}), we search for stationary solutions in the modal space for equations (\ref{final_eq_4dof}). The the derivatives are nullified, and exponential representation for the stationary points $\Psi_{j,0}$ is used:
\begin{equation}
\begin{split}
\Psi_{j,0} = A_j \exp({i \Delta_j})
\end{split}
\end{equation}
Omitting the derivatives and substituting into equations (\ref{final_eq_4dof}), we obtain the following algebraic equation: \\
\begin{align}
\begin{split}
\label{alg_4_DOF}
&2 \Bigg[ 4A_0 - A_0^3 - 2A_0 \left( A_2^2 + A_3^2 \right) -A_2 A_3^2 \exp \left(i\Gamma \right) \Bigg] + \eta \Bigg[ A_0 \Bigg( 3 \left( A_2^4 + A_3^4 \right)\\ 
& + 12A_2^2A_3^2 + 6 A_0^2 \left( A_2^2 + A_3^2 \right) + A_0^4 \Bigg) + A_2A_3^2 \left( \left( 6A_0^2 + 3A_2^2 + 2A_3^2 \right) \exp \left(i\Gamma \right) \right.
\\ 
&\left. + 3A_0 ^2 \exp \left(-i\Gamma \right) \right) \Bigg] = 0 \\
&2 \Bigg[ 4A_2 - A_2^3 - 2A_2 \left( A_0^2 + A_3^2 \right) - A_0 A_3 ^2 \exp \left(i\Gamma \right) \Bigg] 
+ \eta \Bigg[ A_2 \Bigg( 3 \left( A_0^4 + A_3^4 \right) \\& + 12A_0^2A_3^2 
+ 6 A_2 ^2  \left( A_0^2 + A_3^2 \right) + A_2^4 \Bigg)+ A_0A_3^2 \left( \left( 3A_0^2 + 6A_2^2 + 2A_3^2 \right) \exp \left(i\Gamma \right) \right. 
\\
&\left. + 3A_0^2 \exp \left(-i\Gamma \right) \right) \Bigg] = 0 \\
&2 \Bigg[ 4A_3 - A_3^3 - 2A_3 \left( A_0^2 + A_2^2 \right) - A_0 A_2 A_3 \exp \left(-i\Gamma \right) \Bigg] 
+ \eta \Bigg[ A_3 \Bigg( 3 \left( A_0^4 + A_2^4 \right) \\&
+ 12A_0^2A_2^2 + 6 A_3^2 \left( A_0^2 + A_2^2 \right) + A_3^4 \Bigg)+ 6A_0A_2A_3 \left( A_0^2 + A_2^2 + A_3 ^2 \right) + \exp \left(-i\Gamma \right) \\
&  + 2A_0A_2A_3^3 \exp \left(i\Gamma \right) \Bigg] = 0
\end{split}
\end{align}
In system (\ref{alg_4_DOF}), we use the notation $\Gamma = 2\Delta_3 - \Delta_2 - \Delta_0$. The appearance of exponential multipliers clarifies the meaning of resonant terms - the stationary values of the averaged slow-modulation envelopes are achieved under certain "locking conditions" for the phases of the complex variables. From the imaginary parts of the algebraic equations (\ref{alg_4_DOF}) the "locking condition" is derived as follows:
\begin{equation}
\begin{split}
\sin(\Gamma) = 0 \to \Gamma = 0, \pi
\end{split}
\end{equation}
Substituting the solution into the real part of equation (\ref{alg_4_DOF}) yields two sets of algebraic equations for the modal amplitudes:\\
\begin{align}
\begin{split}
&2\Bigg[ 4A_0 - A_0^3 - 2A_0 \left( A_2^2 + A_3^2 \right) \mp A_2 A_3^2 \Bigg] 
\\ &+\eta \Bigg[ A_0 \Bigg( 3 \left( A_2^4 + A_3^4 \right) + 12A_2^2A_3^2 + 6 A_0^2 \left( A_2^2 + A_3^2 \right) + A_0^4 \Bigg)
\\& \pm A_2A_3^2 \left( 9 A_0 ^2 + 3 A_2 ^2 +2 A_3 ^2 \right) \Bigg] = 0 \\
&2 \Bigg[ 4A_2 - A_2^3 - 2A_2 \left( A_0^2 + A_3^2 \right) \pm A_0 A_3 ^2 \Bigg] 
\\ 
&+ \eta \Bigg[ A_2 \Bigg( 3 \left( A_0^4 + A_3^4 \right) + 12A_0^2A_3^2 + 6 A_2 ^2  \left( A_0^2 + A_3^2 \right) + A_2^4 \Bigg)
\\& \pm  A_0 A_3^2 \left( 3 A_0 ^2 + 9 A_2 ^2 + 2 A_3 ^2\right) \Bigg] = 0 \\
&2 \Bigg[ 4A_3 - A_3^3 - 2A_3 \left( A_0^2 + A_2^2 \right) \mp A_0 A_2 A_3 \Bigg] 
\\
& + \eta \Bigg[ A_3 \Bigg( 3 \left( A_0^4 + A_2^4 \right) + 12A_0^2A_2^2 + 6 A_3^2 \left( A_0^2 + A_2^2 \right) + A_3^4 \Bigg)
\\& \pm 2A_0A_2A_3 \left(3A_0^2 + 3A_2^2 + 5 A_3^2 \right) \Bigg] = 0
\end{split}
\end{align}
Any solution of the system allows predicting the shape of the modulation envelope. For example, for $\varphi_1(\tau)$ according to (\ref{3mod_def_1}):\\
\begin{align}
\begin{split}
\label{envelope_4dof_eq}
&\varphi_1 \approx \Psi_0 + \Psi_2 \exp \left(2i \tau \right) +\Psi_2 \exp \left(i \tau \right)= \\
&= \exp (i \Delta_0)\Bigg(A_0 + A_2 \exp(i(2\tau +\Delta_2 + \Delta_0)) + A_3 \exp(i(\tau + \Delta_3 - \Delta_0))\Bigg) \\
&=\exp (i \Delta_0)\Bigg(A_0 + A_2 \exp(i(2\tau ^* - \Gamma)) + A_3 \exp(i \tau^*)\Bigg), \tau ^* = \tau + \Delta_3 - \Delta_0 \\
&|\varphi_1| = \sqrt{A_0 ^2 A_2 ^2 + A_3^2 + 2 A_0 A_3 \cos(\tau^*) \pm 2 A_2 (A_0 \cos(2\tau ^*) + A_3 \cos(\tau ^*)}
\end{split}
\end{align}
The equation with the upper sign corresponds to $\Gamma = 0$, while the equation with the lower sign corresponds to $\Gamma = \pi$. The solutions for $\eta = 0.1$ are as follows:
\begin{align}
\begin{split}
\label{alg_sol}
&\Gamma = 0: A_0 = 1.178313157, A_2 =  1.178313157, A_3 = 1.499773607
\\
&\Gamma = \pi: A_0 = 2.559644497, A_2 =  0.7293707908, A_3 = 1.830273706
\end{split}
\end{align}
In the following figures, the numerical results are compared with the analytical approximations:
\begin{figure}[H]
    \centering
    \includegraphics[width=0.7\textwidth]{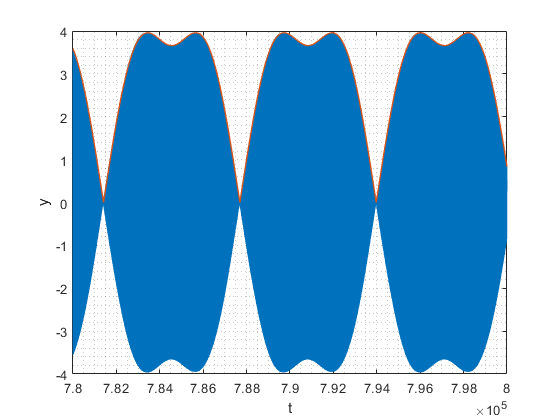}
    \caption{The blue line stands for the numerical simulation for (\ref{in_cond_1}) as the initial conditions, with the parameters $\eta = 0.1, \delta = 0.1, \varepsilon = 0.001$ and the red line stands for the analytical approximation (\ref{envelope_4dof_eq}) with the solution (\ref{alg_sol}).}
    \label{4_dof_verf_hump}
    \includegraphics[width=0.7\textwidth]{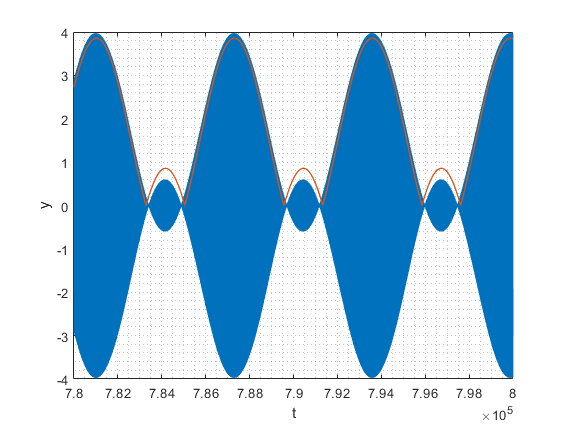}
    \caption{The blue line stands for the numerical simulation for the same initial conditions as in Figure (\ref{4_dof_dbl}), with the parameters $\eta = 0.1, \delta = 0.1, \varepsilon = 0.001$ and the red line stands for the analytical approximation (\ref{envelope_4dof_eq}) with the solution (\ref{alg_sol}).}
    \label{4_dof_verf_dbl}
\end{figure}
As can be seen, there is a good match between the numerical results and the approximation in Figure (\ref{4_dof_verf_hump}), unlike in Figure (\ref{4_dof_verf_dbl}).
\section{Conclusions and discussion}

To our opinion, most important conclusion to be driven from the results presented above is that seemingly somewhat exotic regimes of the "non-conventional" or "modal" synchronization turn generic and ubiquitous if the system is explored in appropriate parametric regime. The reasoning and the treatment is somewhat similar to a simple textbook derivation of the amplitude of VdP limit cycle in the case of small self-excitation: the latter term just "selects" the proper oscillation amplitude, at which the energy balance between positive and negative damping over the cycle is achieved. Current paper demonstrates the validity of similar reasoning for the case of multiple degrees of freedom with generic quasiperiodic foliation of the state space in the limit of zero self-excitation. Then, again, some selected quasiperiodic tori convert to attractors.

It does not seem that there is anything special in our selection of the coupling or of the excitation mechanism. Indeed, somewhat similar beat waves were recently demonstrated in very different setting, with strongly nonlinear coupling and parametric excitation \cite{YS1, YS2}. In the same time, it is possible to say that the models considered in this paper oppose to some extent more conventional approaches to the synchronization. Most traditional models (e.g. Kuramoto model, \cite{pikovsky2001synchronization}, and its generalizations) consider phase - only oscillators. One can claim that the self - excitation in these systems is so strong that the coupling can affect only the phase of oscillations. In this sense, the phenomena of the modal and wave synchronization correspond to the opposite asymptotic regime - the coupling is substantially stronger than the self-oscillation. As it often happens, this alternative asymptotic regime gives rise to multiple phenomena absent in more common models.

The dynamical regimes explored in the paper leave one with substantial challenges and unresolved problems. To mention the most basic ones, it is not clear how one can assess the stability of quasiperiodic attractor, beyond purely numeric approaches. Besides, one can expect that the inhomogeneities, as well as alternative coupling patterns,  will strongly affect the wave synchronization. These and other open questions leave a lot of space for further research.

\section{Declaration of generative AI and AI-assisted technologies in the writing process}
During the preparation of this work, the authors used "ChatGPT" to improve the language quality and grammar. After using this tool/service, the authors reviewed and edited the content as needed and takes full responsibility for the content of the published article.

\section{Acknolegdements}
The authors are very grateful to the Israel Science Foundation (grant 2598/21) for financial support.

\newpage
\appendix
\section{Example Appendix Section}
\label{app1}
The modal equations for the 3 DOF system are as follows:
\small
\begin{align}
\begin{split}
\label{modal_3dof}
\mu_0' &= -\frac{\delta}{16} \Bigg( 2 \Bigg[ 4 \mu_0 - 2 \mu_0^* \mu_1 \mu_2 - \mu_1^2 \mu_2^* - \mu_1^* \mu_2^2 - \mu_0^2 \mu_0^* - 2 \mu_0 \left( \mu_1 \mu_1^* + \mu_2 \mu_2^* \right) \Bigg] \\
&+ \eta \Bigg[ 3 \mu_0 \left( \mu_1^2 \mu_1^{* ^2} + \mu_2^2 \mu_2^{* ^2} \right) + 12 \mu_0 \mu_1 \mu_1^* \mu_2 \mu_2^* + \mu_1 \mu_1^* \left( 6 \mu_0^* \mu_1 \mu_2 + 2 \mu_1^{* ^2} \mu_2^* + 3 \mu_1^* \mu_2^2 + 6 \mu_0^2 \mu_0^* \right) \\
&+ \mu_2 \mu_2^* \left( 6 \mu_0^* \mu_1 \mu_2 + 3 \mu_1^2 \mu_2 + 2 \mu_1^* \mu_2^2 + 6 \mu_0^2 \mu_0^* \right)+ 6 \mu_0 \mu_0^* \left(\mu_0^* \mu_1 \mu_2 + \mu_1^2 \mu_2^* +\mu_1^* \mu_2 ^2\right) \\
&+ 3 \mu_0^2 \mu_1^* \mu_2 + 3\mu_0^2 \mu_1 \mu_2 ^{*^2} + 2\mu_0 ^3 \mu_1^* \mu_2^* +\mu_0^* \left( \mu_1^3 + \mu_2^3  \right) + \mu_0^3 \mu_0^{*^2} \Bigg] \Bigg)\\
\mu_1' &= \frac{3i}{2} \mu_1 - \frac{\delta}{16} \Bigg( 2 \Bigg[ 4 \mu_1 - 2 \mu_0 \mu_1^* \mu_2 - \mu_0^2 \mu_2^* - \mu_0^* \mu_2^2 - \mu_1^2 \mu_1^* - 2 \mu_1 \left( \mu_1 \mu_1^* + \mu_2 \mu_2^* \right) \Bigg] \\
&+ \eta \Bigg[ 3 \mu_2 \left( \mu_0^2 \mu_0^{*^2} + \mu_2^2 \mu_2^{*^2} \right) + 12 \mu_1 \mu_0 \mu_0^* \mu_2 \mu_2^* + \mu_0 \mu_0^* \left( 3 \mu_0^* \mu_2 ^2 +2\mu_0^2 \mu_2^* + 6\mu_0 \mu_1 ^* \mu_2 \right) \\
&+\mu_2 \mu_2^* \left( 2 \mu_0^* \mu_2^2 + 3 \mu_0^2 \mu_2^* + 6 \mu_0 \mu_1^* \mu_2 \right) + \mu_1^{* ^2} (\mu_0 ^3 + \mu_2 ^3) + \mu_1^3 \mu_1^{*^2} + 3\mu_0 \mu_1 ^2 \mu_2 ^{* ^2} \\
&+ 6 \mu_1 \mu_1^* \left( \mu_0 \mu_1^* \mu_2 + \mu_0^2 \mu_2^* + \mu_0^* \mu_2^2 +\mu_1 (\mu_0 \mu_0^* + \mu_2 \mu_2^*) \right) + 3 \mu_0^{*^2} \mu_1^2 \mu_2 + 2 \mu_0^* \mu_1^3 \mu_2^* \Bigg] \Bigg)\\
\mu_2' &= \frac{3i}{2} \mu_2 - \frac{\delta}{16} \Bigg( 2 \Bigg[ 4 \mu_2 - 2 \mu_0 \mu_1 \mu_2^* - \mu_0^* \mu_1^2 -\mu_0^* \mu_1^2 - \mu_2^2 \mu_2^* - 2 \mu_2 \left( \mu_0 \mu_0^*  + \mu_1 \mu_1^* \right) \Bigg] \\
&+ \eta \Bigg[ 3 \mu_2 \left( \mu_0^2 \mu_0^{*^2} + \mu_1^2 \mu_1^{*^2} \right) + 12 \mu_2 \mu_0 \mu_0^* \mu_1\mu_1^* + \mu_0 \mu_0^* \left( 3\mu_0 ^* \mu_1 ^2 + 2 \mu_0^2 \mu_1^* +6\mu_0 \mu_1 \mu_2 ^* \right) \\
&+ \mu_1 \mu_1^* \left( 2 \mu_0^2 \mu_1^2 + 3 \mu_0^2 \mu_1^* + 6 \mu_0 \mu_1 \mu_2^* \right) + \mu_2^{* ^2} \left( \mu_0^3 + \mu_1^3 \right) + \mu_2^3 \mu_2^* + 3 \mu_0 \mu_1^{* ^2} \mu_2 ^2 \\
&+ 6 \mu_2 \mu_2^* \left( \mu_0 \mu_1 \mu_2^* + \mu_0^2 \mu_1^* + \mu_0^* \mu_1^2 +\mu_2 (\mu_0 \mu_0^* + \mu_1 \mu_1 ^*) \right) + 3 \mu_0^{*^2} \mu_1 \mu_2^2 + 2 \mu_0^* \mu_1 ^* \mu_2 ^3 \Bigg] \Bigg)
\end{split}
\end{align}
\newpage
The modal equations for the 3 DOF system after substitution of (\ref{rescale_eq_1}) are as follows:
\footnotesize
\begin{align}
\begin{split}
\label{exp_3dof}
\frac{d\psi_0}{d\xi} = & -\frac{1}{16} \Bigg( 2 \left[ 4 \psi_0 - 2 \psi_0^* \psi_1 \psi_2 \exp\left(\frac{3i \xi}{\delta}\right) - (\psi_1 ^2 \psi_2 ^* + \psi_1^* \psi_2^2) \exp\left(\frac{3i \xi}{2\delta}\right) - \psi_0^2 \psi_0^*  -2 \psi_0 \left( \psi_1 \psi_1^* + \psi_2 \psi_2^* \right) \right] \\
& + \psi_1 \psi_1^* \left( 6 \psi_0^* \psi_1 \psi_2 \exp \left(\frac{3i \xi} {\delta} \right) + (2 \psi_1 ^2 \psi_2 ^* + 3 \psi_1^* \psi_2 ^2) \exp\left( \frac{3i \xi} {2 \delta} \right) + 6 \psi_0^2 \psi_0^* \right) \\
& + \psi_2 \psi_2 ^* \left( 6 \psi_0^* \psi_1 \psi_2 \exp \left(\frac{3i \xi}{\delta} \right) + (3 \psi_1^2 \psi_2^* + 2 \psi_1^* \psi_2 ^2) \exp \left(\frac{3i \xi}{2 \delta} \right) + 6 \psi_0^2 \psi_0 ^* \right) \\
& + 6  \psi_0 \psi_0^*  \left( \psi_0^* \psi_1 \psi_2 \exp \left(\frac{3i \xi}{\delta} \right) + (\psi_1^2 \psi_2 ^* + \psi_1^* \psi_2 ^2) \exp \left( \frac{3i \xi}{2 \delta} \right) \right) \\
& + 3 \left( \psi_0 ^2  \psi_1 ^{* ^2} \psi_2 + \psi_0 ^2 \psi_1 \psi_2 {*^2} \right) \exp \left(\frac{-3i \xi}{2 \delta}\right) + 2 \psi_0 ^3 \psi_1 ^* \psi_2 ^*  \exp \left(\frac{-3i \xi}{ \delta} \right) \\
& + \psi_0 ^{*^2} \left( \psi_1 ^3 + \psi_2 ^3 \right) exp \left( \frac{9i \xi}{2 \delta} \right)  + \psi_0^3 \psi_0 ^{*^2} + 12 \psi_0 \psi_1 \psi_1 ^* \psi_2 \psi_2 ^* + 3 \psi_0 \left( \psi_1 ^2 \psi_1 {*^2} +\psi_2 ^2 \psi_2 {*^2}\right) \Bigg) \\
\frac{d\psi_1}{d\xi} = & -\frac{\delta}{16} \Bigg[ 2 \Bigg( 4 \psi_1 - 2 \psi_0 \psi_1 ^* \psi_2 \exp\left(\frac{-3i \xi}{2 \delta}\right) - \psi_0^2 \psi_2 ^* \exp\left(\frac{-3i \xi} {\delta}\right) - \psi_0^*  \psi_2 ^2  \exp \left(\frac{3i \xi}{2 \delta} \right)\\
& -\psi_1 ^2 \psi_1 ^* -2 \psi_1 \left(\psi_0 \psi_0 ^* + \psi_2 \psi_2 ^* \right) \Bigg)   \\
& +\eta \Bigg( \psi_0 \psi_0^* \left( 6 \psi_0 \psi_1 ^* \psi_2 \exp \left(\frac{-3i \xi} {2 \delta} \right) + 2 \psi_0 ^2 \psi_2 ^* \exp \left(\frac{-3i \xi}{\delta} \right) + 3 \psi_0  \psi_2 ^{*^2} \exp \left(\frac{3i \xi} {\delta}\right) \right) \\
& + \psi_2 \psi_2 ^* \left( 6 \psi_0 \psi_1 ^* \psi_2 \exp \left(\frac{3i \xi} {2 \delta} \right) + 3 \psi_0 ^2 \psi_2 ^* \exp\left(\frac{-3i\xi}{\delta} \right) + 2 \psi_0 ^* \psi_2 ^2 \exp\left(\frac{3i\xi}{2\delta} \right) \right)  \\
& + \psi_1 ^{*^2} \left(\psi_0 ^3 \exp \right(\frac{9i \xi}{2\delta} \left) + \psi_2 ^3 \right) + \psi_1 ^3 \psi_1 ^{*^2} + 3\psi_0 \psi_1^2 \psi_2 ^{*^2} \exp \left(\frac{-3i\xi}{2\delta} \right) \\
& + 6 \psi_1 \psi_1^* \left( \psi_0 \psi_1^* \psi_2 \exp \left(\frac{-3i \xi}{2 \delta} \right) + \psi_0^2 \psi_2 ^*  \exp \left(\frac{-3i \xi}{2 \delta} \right) + \psi_0^* \psi_2 ^2  \exp \left(\frac{3i \xi}{2 \delta} \right) + \psi_1 (\psi_0 \psi_0 ^* + \psi_2 \psi_2^*)\right) \\
& + 3\psi_1 \left( \psi_0^2 \psi_0 ^{*^2} + \psi_1^2 \psi_1 ^{*^2} \right) + 12 \psi_1 \psi_0 \psi_0 ^* \psi_2 \psi_2 ^* + 3\psi_0 ^{*^2} \psi_1^2 \psi_2 \exp \left( \frac{3i\xi}{\delta} \right) + 2\psi_0 ^* \psi_1 ^3 \psi_2 ^* \exp \left( \frac{3i \xi}{2\delta} \right) \Bigg)  \Bigg] \\
\frac{d\psi_2}{d\xi} = & -\frac{\delta}{16} \Bigg[ 2 \Bigg( 4 \psi_2 - 2 \psi_0 \psi_1 \psi_2^* \exp \left(\frac{-3i \xi} {2 \delta} \right) - \psi_0 \psi_1 ^* \exp \left(\frac{-3i \xi} {\delta} \right) - \psi_0 ^* \psi_1 ^2 \exp \left(\frac{3i \xi} {\delta} \right) \\
& -\psi_2 ^3 \psi_2 ^{*^2} - 2\psi_2 \left( \psi_0 \psi_0 ^* + \psi_1 \psi_1^* \right)\Bigg) \\
& +\eta \Bigg( \psi_0 \psi_0^* \left( 6 \psi_0 \psi_1 \psi_2^* \exp \left(\frac{-3i \xi} {2 \delta} \right) + 2 \psi_0 ^2  \psi_1^* \exp \left(\frac{-3i \xi} {\delta} \right) + 3 \psi_0 ^* \psi_1 ^2  \exp \left(\frac{3i \xi} {2\delta} \right) \right) \\
& + \psi_1 \psi_1 ^*  \left( 6 \psi_0 \psi_1 \psi_2 ^* \exp \left( \frac{- 3i \xi} {2 \delta} \right) +3 \psi_0 ^2 \psi_1 ^*  \exp \left(\frac{-3i \xi}{\delta} \right) + 2 \psi_0 ^* \psi_1 ^2  \exp \left(\frac{3i \xi} { 2 \delta} \right) \right) \\
& \psi_2 ^{*^2} \left( \psi_0 ^3 \exp \left( \frac{9i\xi}{2\delta} \right) + \psi_1 ^3 \right) + \psi_2 ^3 \psi_2 ^2 +3\psi_0 \psi_1 ^{*^2} \psi_2 ^2 \exp \left(\frac{-3i\xi}{\delta} \right)  \\
& + 6 \psi_2 \psi_2 ^* \left( \psi_0 \psi_1 \psi_2 ^* \exp \left( \frac{-3i\xi}{2\delta} \right) + \psi_0 ^2 \psi_1 ^* \exp \left( \frac{-3i\xi}{\delta} \right)
+ \psi_0^* \psi_1 ^2 \exp \left( \frac{3 i \xi}{2 \delta} \right) + \psi_2 \left( \psi_0 \psi_0 ^* + \psi_1 \psi_1 ^*  \right)  \right) \\
&  + 3 \psi_2 \left(\psi_0 ^2 \psi_0 ^{*^2} + \psi_1 ^2 \psi_1 ^{*^2} \right) + 12 \psi_2 \psi_0 \psi_0 ^* \psi_1 \psi_1 ^* + 3\psi_0 ^{*^2} \psi_1 \psi_2 ^2 \exp \left( \frac{3 i \xi}{\delta} \right) + 2\psi_0 ^* \psi_1 ^* \psi_2 ^3 \exp \left(\frac{3 i \xi}{2 \delta}\right)  \Bigg) \Bigg]
\end{split}
\end{align}
\newpage
\normalsize
The modal equations for the 4 DOF system after substitution of (\ref{rescale_eq_1}) are as follows:
\footnotesize
\begin{equation}
\begin{split}
\frac{d\Psi_0}{d\xi} & = -\frac{1}{16} \Bigg( 2 \Bigg[ 4 \Psi_0 - \Psi_0 ^2 \Psi_0 ^* - 2\Psi_0 \left(\Psi_1 \Psi_1 ^*  + \Psi_2 \Psi_2 ^* + \Psi_3 \Psi_3 ^* \right) - \left(\Psi_1^2 + \Psi_3 ^2 \right) \Psi_2 ^* \Bigg] \\
&+ \eta \Bigg[ \Psi_0 \Bigg( 3 \left(\Psi_1 ^2 \Psi_1 ^{*^2} + \Psi_2 ^2 \Psi_2 ^{*^2} + \Psi_3 ^2 \Psi_3 ^{*^2} \right) + 12 \left( \Psi_1 \Psi_1 ^* \Psi_2 \Psi_2 ^* + \Psi_1 \Psi_1 ^* \Psi_3 \Psi_3 ^* + \Psi_2 \Psi_2 ^* \Psi_3 \Psi_3 ^* \right) \\
&+ 6 \Psi_0 \Psi_0 ^* \left( \Psi_1 \Psi_1 ^*  + \Psi_2 \Psi_2 ^* + \Psi_3 \Psi_3 ^*  \right) + \Psi_0 ^2 \Psi_0 ^ {*^2} \Bigg) + 3 \Psi_1 ^{*^2} \left( \Psi_0 ^2 \Psi_2 + \Psi_0 \Psi_3 ^2 \right) \\
&+ \Psi_2 ^* \Bigg( 6 \Psi_0 \Psi_0 ^* \left( \Psi_1 ^2 + \Psi_3 ^2 \right) + 2 \Psi_1 \Psi_1 ^* \left( \Psi_1 ^2 + 3 \Psi_3 ^2  \right) + 3 \Psi_2 \Psi_2 ^ * \left( \Psi_1 ^2 + \Psi_3 ^2 \right) \\
& + 2 \Psi_3 \Psi_3 ^* \left(3 \Psi_1 ^2 + \Psi_3 ^2 \right) \Bigg) + 3 \Psi_3 ^{*^2} \left( \Psi_0 ^2 \Psi_2 + \Psi_0 \Psi_1 ^2 \right) \Bigg] \Bigg) \\
\frac{d\Psi_1}{d\xi} &= -\frac{1}{16} \Bigg( 2 \Bigg[ 4 \Psi_1 - \Psi_1 ^2  \Psi_1 ^* - 2\Psi_1 \left(\Psi_0 \Psi_0 ^*  + \Psi_2 \Psi_2 ^* + \Psi_3 \Psi_3 ^* \right) - \left(2 \Psi_0 \Psi_2 + \Psi_3 ^2 \right) \Psi_1 ^* \Bigg] \\
&+ \eta \Bigg[ \Psi_1 \Bigg( 3 \left(\Psi_0 ^2 \Psi_0 ^{*^2} + \Psi_2 ^2 \Psi_2 ^{*^2} + \Psi_3 ^2 \Psi_3 ^{*^2} \right) + 12 \left( \Psi_0 \Psi_0 ^* \Psi_2 \Psi_2 ^* + \Psi_0 \Psi_0 ^* \Psi_3 \Psi_3 ^* + \Psi_2 \Psi_2 ^* \Psi_3 \Psi_3 ^* \right) \\
&+ 6 \Psi_1 \Psi_1 ^* \left( \Psi_0 \Psi_0 ^*  + \Psi_2 \Psi_2 ^* + \Psi_3 \Psi_3 ^*  \right) + \Psi_1 ^2 \Psi_1 ^{*^2} \Bigg) + \Psi_1 ^* \Bigg( 6 \Psi_0 \Psi_0 ^* \left( \Psi_0 \Psi_2 + \Psi_3 ^2 \right) \\
& + 3 \Psi_1 \Psi_1 ^* \left( 2 \Psi_0 \Psi_2 + \Psi_3 ^2 \right) + 6 \Psi_2 \Psi_2 ^* \left( \Psi_0 \Psi_2 + \Psi_3 ^2 \right) + 2 \Psi_3 \Psi_3 ^* \left( 6 \Psi_0 \Psi_2 + \Psi_3 ^2 \right) \Bigg) \\
&+ \Psi_1 \Psi_3 ^{*^2} \left( 6 \Psi_0 \Psi_2 + \Psi_1 ^2  \right) + 2 \Psi_0 ^* \Psi_1 \Psi_2 ^* \left(\Psi_1 ^2 + 3 \Psi_3 ^2 \right) \Bigg] \Bigg) \\
\frac{d\Psi_2}{d\xi} &= -\frac{1}{16} \Bigg( 2 \Bigg[ 4 \Psi_2 - \Psi_2 ^2  \Psi_2 ^* - 2\Psi_2 \left(\Psi_0 \Psi_0 ^*  + \Psi_1 \Psi_1 ^* + \Psi_3 \Psi_3 ^* \right) - \left( \Psi_1 ^2 + \Psi_3 ^2 \right) \Psi_0 ^* \Bigg] \\
&+ \eta \Bigg[ \Psi_2 \Bigg( 3 \left(\Psi_0 ^2 \Psi_0 ^{*^2} + \Psi_1 ^2 \Psi_1 ^{*^2} + \Psi_3 ^2 \Psi_3 ^{*^2} \right) + 12 \left( \Psi_0 \Psi_0 ^* \Psi_1 \Psi_1 ^* + \Psi_1 \Psi_1 ^* \Psi_3 \Psi_3 ^* + \Psi_0 \Psi_0 ^* \Psi_3 \Psi_3 ^* \right) \\
&+ 6 \Psi_2 \Psi_2 ^* \left( \Psi_0 \Psi_0 ^*  + \Psi_1 \Psi_1 ^* + \Psi_3 \Psi_3 ^*  \right) + \Psi_2 ^2 \Psi_2 ^{*^2} \Bigg) + 3 \Psi_1 ^{*^2} \left( \Psi_0 \Psi_2 ^2 + \Psi_2 \Psi_3 ^2 \right) + \Psi_0 ^* \Bigg( 3 \Psi_0 \Psi_0 ^* \left( \Psi_1 ^2  + \Psi_3 ^2 \right) \\
& + 2 \Psi_1 \Psi_1 ^* \left( \Psi_1 ^2 + 3 \Psi_3 ^2 \right) + 6 \Psi_2 \Psi_2 ^* \left( \Psi_1 ^2 + \Psi_3 ^2 \right) + 2 \Psi_3 \Psi_3 ^* \left( 3 \Psi_1 ^2 + \Psi_3 ^2 \right) \Bigg) + 3 \Psi_3 ^ {*^2} \left( \Psi_0 \Psi_2 ^2 + \Psi_2 \Psi_1 ^2  \right) \Bigg] \Bigg) \\
\frac{d\Psi_3}{d\xi} &= -\frac{1}{16} \Bigg( 2 \Bigg[ 4 \Psi_3 - \Psi_3 ^2  \Psi_3 ^* - 2\Psi_3 \left(\Psi_0 \Psi_0 ^*  + \Psi_1 \Psi_1 ^* + \Psi_2 \Psi_2 ^* \right) - \left(2 \Psi_0 \Psi_2 + \Psi_1 ^2 \right) \Psi_3 ^* \Bigg] \\
&+ \eta \Bigg[ \Psi_3 \Bigg( 3 \left(\Psi_0 ^2 \Psi_0 ^{*^2} + \Psi_1 ^2 \Psi_1 ^{*^2} + \Psi_2 ^2 \Psi_2 ^{*^2} \right) + 12 \left( \Psi_0 \Psi_0 ^* \Psi_2 \Psi_2 ^* + \Psi_0 \Psi_0 ^* \Psi_1 \Psi_1 ^* + \Psi_1 \Psi_1 ^* \Psi_2 \Psi_2 ^* \right) \\
&+ 6 \Psi_3 \Psi_3 ^* \left( \Psi_0 \Psi_0 ^*  + \Psi_1 \Psi_1 ^* + \Psi_2 \Psi_2 ^*  \right) + \Psi_3 ^2 \Psi_3 ^{*^2} \Bigg) + \Psi_3 ^* \Bigg( 6 \Psi_0 \Psi_0 ^* \left( \Psi_0 \Psi_2 + \Psi_1^2  \right) + 2 \Psi_1 \Psi_1 ^* \left( 6 \Psi_0 \Psi_2 + \Psi_1 ^2 \right)\\
& 6 \Psi_2 \Psi_2 ^*  \left( \Psi_0 \Psi_2 + \Psi_1 ^2 \right) + 3 \Psi_3 \Psi_3 ^* \left( 2 \Psi_0 \Psi_2 + \Psi_1 ^2 \right) \Bigg) + \Psi_1 ^{*^2} \Psi_3 \left( 6 \Psi_0 \Psi_2 + \Psi_3 ^2 \right) + 2 \Psi_0 ^* \Psi_3 \Psi_2 ^* \left( 3 \Psi_1 ^2 + \Psi_3 ^2 \right)\Bigg] \Bigg) \\
\end{split}
\end{equation}
\newpage
\bibliographystyle{elsarticle-num-names} 
\bibliography{YW_OG_refrences}






\end{document}